\documentclass[12pt,a4paper]{article}

 \renewcommand{\baselinestretch}{1.4}

\usepackage{amsmath}
\numberwithin{equation}{section} 

 \newenvironment{myabstract}[1]{
 \renewcommand{\baselinestretch}{1.4}
 \vspace*{10mm} \large
 \begin{center} {\bf Abstract} \end{center}
 \begin{center} \parbox{14cm}{\small #1}}{\end{center}}

\begin{document}

 \renewcommand{\baselinestretch}{1.7}

 \title{\vspace{-20mm} Numerical correspondences between the physical constants}

 \author{Christophe R\'{e}al \\ 22, rue de Pontoise, 75005 PARIS \vspace{-3mm} \\
 \vspace{4mm}}

\date{}

 \maketitle

\vspace{10mm}

 \begin{myabstract}{

 We present here a note which synthesizes our previous ideas
 concerning some problems in cosmology, and the numerical
 correspondences between the constants in physics which we could
 deduce.

 }\end{myabstract}

\vspace{10mm}

\section{Tensorial quantum gravity}
\subsection{The equation}
 In $[1]$, we consider Einstein's
equations with a cosmological constant :
\begin{equation}R_{ik}-\frac{1}{2}R+\Lambda g_{ik}=\kappa
T_{ik}\end{equation} We know that $\Lambda$ and $\kappa$ are
constants, and the equation corresponds to the well known
Hilbert-Einstein action with cosmological constant. We ask the
question : is there some other equation, constructed from (1.1) by
simple changes, thus still tensorial, which would take into
account some quantum features, beyond classical gravity? Looking
at the other interactions, we see that after renormalization, the
coupling constants are running on energy (perturbative
corrections) and that there also appears nonperturbative
corrections, related to topology (instantons for example). Thus,
we investigate (1.1) in the case in which $\kappa$ is running on
energy density, and for $\Lambda$ being the Gauss-Bonnet
topological term. More precisely we call $\tilde{\Sigma}$ the
Gauss-Bonnet term and define
$\tilde{\Sigma}=4\tilde{\tilde{\Sigma}}$, which will simplify the
calculations : \begin{equation}\tilde{\Sigma} = R^{(4)} -4
R^{(2)}+ R^{2}=4\tilde{\tilde{\Sigma}}\end{equation} where
$R^{(4)}=R^{abcd}R_{abcd}$ and $R^{(2)} =R^{ab}R_{ab}$, $R_{abcd}$
being the Riemann curvature tensor, $R_{ab}$ the Ricci tensor, and
$R$ the scalar curvature. This term is on the left hand side of
our equation and we obtain :
\begin{equation}
R_{ik}-\frac{1}{2}Rg_{ik}+\Lambda g_{ik} =\kappa(\epsilon)T_{ik}
\end{equation}
with :
\begin{equation}
\Lambda =
-\theta\tilde{\tilde{\Sigma}}+\Lambda_{0}=-\frac{\theta\tilde{\Sigma}}{4}+\Lambda_{0}
\end{equation}
In a first case, we suppose $\theta$ constant, and we add a
constant $\Lambda_{0}$. Then, in the general case, we suppose
$\theta$ to be any function, for example $\theta=\theta(a)$, and
we impose $\Lambda_{0}=0$. Equation (1.3) looks tensorial,
classical, and nevertheless the new coupling of gravity has become
dimensionless in one case. Indeed, using (1.3) and (1.4), we can
prove that if the coupling $\kappa(\epsilon)$ takes the form :
\begin{equation}
\kappa(\epsilon)=\frac{\kappa_{1}}{\sqrt{\hbar c}\sqrt{\epsilon}}
\end{equation}
then $\kappa_{1}$ a dimensionless real number. We will still note
this constant $\kappa_{1}$ when considering $\hbar=c=1$.
\subsection{A remark}
Equation (1.3) could appear inconsistent at first sight for the
following reason. The left hand side of (1.1), associated to the
constancy of $\kappa$, implies automatically the conservation of
energy of the matter fields $T_{ik}$, which is not the case in
equation (1.3). Thus, when considering equation (1.3), we also
have to add to these equations the four equations of conservation
of the matter fields. This does not make more equations than
unknowns, because in (1.3) we add all matter fields as unknowns.
In other words, (1.1) is the equation for the only gravitational
field, and (1.3) is the equation for all fields. This explains why
our equation gives the value of $p$, which has not to be put by
hand anymore. Still, there could be a possible inconsistency in
(1.3) : if (1.3) also governs the behavior of the matter fields,
it could be in contradiction with the conservation of entropy. The
present article answers to this question, since we prove here that
in the case of the Robertson-Walker metric, (1.3) implies that the
conservation of entropy is equivalent to the conservation of
energy of the matter fields.
\section{The predictions : constant theta}
We emphasize that the equation (1.3) (associated with (1.4) and
(1.5)), even if it does not come at first sight from any
lagrangian, has very interesting properties concerning the
description of the early universe. First, it possesses one
equation more than the cosmological models based on general
relativity. Indeed, we will see that this equation provides us
with the relation
\begin{equation}p=\epsilon/3\end{equation}
which is not put by hand anymore but can be deduced from (1.3). As
in the cosmological models based on general relativity, entropy is
a consequence of the equations of movement. Above all, as far as
the early universe is concerned (we are necessarily in the case
$p=\epsilon/3$), the equation is a kind of unification of the
standard cosmological model and of the inflationary universe.
Indeed, we have
\begin{equation}\epsilon\sim\frac{1}{a^{4}}\end{equation}
and $p>0$, like in the standard model (these last two equations
are consequences of (2.1) and of the conservation of entropy). But
also, the behavior of $a(t)$ is like in the inflationary universe
:
\begin{equation}a(t)=\exp(Ht)\end{equation} with $H$ constant. So
there is another consequence of (1.3) : it smoothes out the
initial singularity with no need for a scalar particle. We
emphasize that this way is the unique possibility to solve the
problem of the beginning of the universe. The absence of a
beginning cannot be due to a scalar particle belonging to the
universe itself, because logically, the beginning is BEFORE the
existence of the particle, which cannot then determine the absence
of the beginning, by a kind of logical retroaction. We emphasize
also that such an equation which gives $p$, obliges matter to be
relativistic. In conclusion, we arrive in the case of constant
$\theta$ to the conclusion : \emph{one sole tensorial equation of
quantum gravity governs at the same time the cosmological
parameters of the universe, exactly as did the standard
cosmological model, and the structure of the fundamental
particles, giving the right relation between $p$ and $\epsilon$,
whereas, furthermore, it smoothes out the initial singularity with
no further hypothesis.}
\section{The predictions : Varying theta (the general case)}
\subsection{The equations} In this case, the equation of quantum
gravity is equivalent to a system of three equations, which are :
\begin{equation}\frac{\ddot{a}}{a}=\frac{1}{4\theta(a)}\end{equation}
We also have, when (1.5) is used :
\begin{equation}
\frac{K}{a^{2}}+H^{2}=\frac{2}{3}\kappa_{1}\sqrt{\epsilon}
\end{equation}
and for the pressure :
\begin{equation}p=\frac{\epsilon}{3}\left(1-\frac{1}{\frac{2}{3}\kappa_{1}\sqrt{\epsilon}\theta(a)}\right)\end{equation}
\subsection{An accelerating expansion}
Equation (3.3) combined to the fact that we have on $p$ the
constraint :
$$0\leq p\leq\frac{\epsilon}{3}$$ proves that necessarily :
\begin{equation}\theta(a)>0\end{equation}
Using (3.1), we conclude that $\ddot{a}>0$, which means that the
expansion is accelerating. We emphasize that this experimental
feature of our universe is not explained by the usual cosmological
models in vigor, and that finding a model which could explain this
fact has been the subject of intense researches since some time.
This equation also solves the expansion problem since it explains
why our universe is in expansion (because $\ddot{a}$ has ALWAYS
been positive, so $\dot{a}$ has always been increasing, which
explains why it has become positive).
\subsection{The absence of a beginning}
Our function $a$ is convex, and of course positive, so we have two
possibilities. With constant $\theta$ for example, we have an
exponentially growing universe, with no beginning. In fact, we
prove in $[1]$ that we rather are, once taken into account ALL
constraints on our model, in the situation with a universe
shrinking to a minimum $a_{0}$, and expanding again from this
value. We note $t=0$ the cosmological time of the minimum.
\subsection{The closed model}
We necessarily are in the closed model, because at the minimum, or
in any point where $\dot{a}$ is small enough, and since $K=\pm1$,
equation (3.2), which exhibits a positive quantity on the right
hand side, implies $K=+1$ (closed model).
\subsection{A new class of cosmological models}
Equation (3.1) proves that any convex behavior of $a(t)$ can be
reproduced by some positive function $\theta(a)$. Furthermore, we
can change the behavior of $\kappa(\epsilon)$ by any other formula
in which $1/\sqrt{\epsilon}$ of (1.5) is replaced by any function
which has the dimension of a square of an energy. Another, even
preferred choice would be :
\begin{equation}\kappa(\epsilon)=\frac{\kappa_{1}^{2}}{H^{2}}\end{equation}
In all, we have two functions in our equation which can be chosen
almost at will, in order to fit the whole set of cosmological
experiments.
\subsection{A particular choice of theta}
We have studied all the models, only one, and even the choice
(1.5) has not been our physically preferred choice, but the one
which gave the simplest calculations! For a physical choice of
$\theta(a)$, we refer to the results of constant $\theta$, because
in this case, $\sqrt{\theta}$, which is a length, is equal to
$c/2H$. We thought that such a huge length could simply be the
radius of the universe itself (which led us for the first time to
suspect the fact that $\theta$ is in fact varying). Thus we
suspected that $\sqrt{\theta}\approx a$, and we posed :
\begin{equation}\theta(a)=\theta_{0}a^{2}\end{equation}
$\theta_{0}$ being a constant, which will be proved in $[3]$ to be
around unity (with some interesting hypotheses, we can prove that
$2\theta_{0}=1$.
\subsection{Value of the time derivative of a}
Using now (3.6), we can integrate (3.1) and find :
\begin{equation}\dot{a}(t)=\frac{1}{\sqrt{2\theta_{0}}}\sqrt{\ln(a/a_{0})}\end{equation}
Using for the relation $2\theta_{0}=1$, we can compute the maximal
value of the present $\dot{a}(t_{pr})$, by taking for $a_{0}$ the
Planck length, and we find : $\dot{a}(t_{pr})\leq11.7$. For the
minimal value of $a_{pr}=a(t_{pr})$, we can choose the condition
$(a/a_{0})\geq6$, since it has been observed quasars with
redshifts around $z=5$. This leads to
$$1.34\leq\dot{a}(t_{pr})\leq11.7$$
For a very probable value inside this interval, we choose for
$a_{0}$ a value less than or equal to the radius of the early
universe in the standard cosmological model ($z=10^{10}$) :
$$\dot{a}(t_{pr})\geq4.8$$
What appears in any case is that our quantum cosmological model
predicts for $\dot{a}$ a present value around unity.
\subsection{The age of the universe} Observations in the context
of the standard cosmology (Bennett, 2003, $[5]$) make appear the
fact that the age of the universe should be very near the inverse
of Hubble's constant. We note $t$ the cosmological time, and it is
a simple mathematical exercise to prove that from the behavior
(3.7), we can prove, independently on the value $\theta_{0}$, that
\begin{equation}t\sim\frac{1}{H}\end{equation}
where we mean, by $f(t)\sim g(t)$, that the ratio of these two
functions tends to $1$ when $t\rightarrow+\infty$.
\subsection{The cosmological constant problem}
\subsubsection{The classical context}
As far as the standard model of cosmology is concerned, the
cosmological parameters of the model are measured with a very good
approximation (Bennett and al. 2003, $[5]$). In particular, there
are in this model two important parameters, the total energy
density $\Omega_{TOT}\approx1$ and the energy density of dark
energy $\Omega_{\Lambda}$, the observed relation being :
\begin{equation}\Omega_{\Lambda}\approx\frac{3}{4}\Omega\end{equation}
There is a lot of dark energy density, which remains unexplained.
Furthermore, the model uses the equation of general relativity,
with a cosmological constant $\Lambda$ :
\begin{equation}R_{ik}-\frac{1}{2}Rg_{ik}-\Lambda g_{ik}=8\pi GT_{ik}\end{equation}
Now the term $\Omega_{\Lambda}$ is defined by the formula :
\begin{equation}\Omega_{\Lambda}=\frac{\Lambda}{3H^{2}}\approx\frac{3}{4}\end{equation}
The cosmological constant problem is to understand why a constant
like $\Lambda$ should be nonzero, and furthermore should possess
such a tiny strictly positive value :
\begin{equation}\Lambda\approx\frac{9H^{2}}{4}\end{equation}
Finally, we can state the problem in the following way : the
equations of general relativity are in the number of two, one
which gives $\epsilon$, the other gives $p$. We have, furthermore,
the equation of conservation of entropy, so three equations plus
the fact that they are dependent. So we choose two equations, say
the conservation of entropy and :
\begin{equation}R_{0}^{0}-\frac{1}{2}R-\Lambda =8\pi G\epsilon\end{equation}
If we pass the constant $\Lambda$ to the right hand side of the
equation, and insert it in the term in $\epsilon$, we find a new
$\epsilon$, which we could call $\epsilon_{app}$, because it is an
apparent energy density. The observed value of $\Lambda$ in the
classical context is such that :
\begin{equation}\epsilon_{app}=4\epsilon\end{equation}
As a remark, $\epsilon_{app}$ is the value of the observed energy
density, when the equation without the cosmological term is used,
that is to say we have
$$R_{0}^{0}-\frac{1}{2}R=8\pi G\epsilon_{app}$$
The other equation just gives the relation between $p$ and
$\epsilon$. If we want this conservation of entropy still to be
valid for apparent quantities, we have to pose $$p_{app}=4p$$
Since in the case of the standard model we have $p=0$, this does
not change anything for the value of the pressure.
\subsubsection{The quantum context}
In the context of the quantum equation, we know the origin of the
$\Lambda$ term : it is the Gauss-Bonnet term which we have
inserted in order to take into account some kind of
nonperturbative (topological) quantum corrections to classical
gravity. The computations made in $[1]$ prove that we have, for
$G_{0}^{0}=R_{0}^{0}-\frac{1}{2}R$ :
\begin{equation}G_{0}^{0}=3\left(\frac{K}{a^{2}}+H^{2}\right)\end{equation}
In the quantum context, we also have two equations, plus the
conservation of entropy, and they also are dependent. We can
choose the equation ($[1]$) :
\begin{equation}G_{0}^{0}+\Lambda=\frac{3}{2}\left(\frac{K}{a^{2}}+H^{2}\right)=\kappa_{1}\sqrt{\epsilon}\end{equation}
and the conservation of entropy. Then, we have to analyze (3.16),
and how $\epsilon$ is affected by forgetting the $\Lambda$
Gauss-Bonnet term. If we forget the $\Lambda$ term in (3.16), we
have to replace
$$G_{0}^{0}+\Lambda=\kappa_{1}\sqrt{\epsilon}$$ by :
\begin{equation}G_{0}^{0}=\kappa_{1}\sqrt{\epsilon_{app}}\end{equation}
where $\epsilon_{app}$ is the apparent matter density, exactly as
we did in our analysis of the case of general relativity. The
difference is that now $\epsilon_{app}$ can be calculated
theoretically from the quantum equations and compared to the
original $\epsilon$. In $[1]$ we prove that :
\begin{equation}\Lambda=-\frac{1}{2}G_{0}^{0}\end{equation}
Forgetting $\Lambda$ in our equation would have the net effect of
changing
$$G_{0}^{0}+\Lambda=\frac{1}{2}G_{0}^{0}$$ for $G_{0}^{0}$. So we see
that the net effect of forgetting the $\Lambda$-term on the left
hand side of the equation is to multiply the right hand side by
$2$, which has the effect of doubling $\kappa_{0}$, if we
interpret this change in terms of a change of the gravitational
constant. However, if we prefer to interpret the change in the
equation as a change in $\epsilon$, we get the striking relation :
\begin{equation}\epsilon_{app}=4\epsilon\end{equation}
This factor $4$ corresponds to a prediction of the quantum
equation in the quantum context, and is equal to the factor $4$
coming from the observations, in the context of general
relativity.
\subsubsection{Complete calculation of the cosmological constant}
If we compute $\Omega_{\Lambda}$ using the fact that our
observations of the values of the masses in the cosmos are only
based on the principle of inertia, that is to say, are based on
what we see of the deformations of spacetime from a flat metric,
we find :
\begin{equation}\Omega_{TOT}=\frac{G_{0}^{0}}{3H^{3}}\end{equation}
We recall that we had (3.18) :
\begin{equation}\Lambda=-\frac{1}{2}G_{0}^{0}\end{equation}
and from (3.15) and (3.16) :
\begin{equation}G_{0}^{0}=3\left(\frac{K}{a^{2}}+H^{2}\right)=2\kappa_{1}\sqrt{\epsilon}\end{equation}
So we obtain :
\begin{equation}\Omega_{TOT}=\frac{2\kappa_{1}\sqrt{\epsilon}}{3H^{2}}\end{equation}
We know that $\Lambda$ is negative because it possesses an extra
minus sign compared to the usual $\Lambda$ of general relativity.
Putting all these relations together, we find that our equation
predicts for the usual $\Lambda$ a positive value, verifying :
\begin{equation}\Omega_{\Lambda}=\frac{\Lambda}{3H^{2}}=\frac{1}{2}\Omega_{TOT}\end{equation}
which is clearly in the domain of uncertainties of the
observations, since this domain is determined by the relations
$$-1<\frac{\Lambda}{3H^{2}}<2$$
With $\Omega=1.02$ (observations of Bennett, 2003, $[5]$), our
$\Lambda$ is just at the center of the former interval.
\subsection{Determining kappa}
\subsubsection{The other kappa}
We now observe that the coefficient $4$ between the true physical
and apparent energy densities is only $4$ because it is viewed
from the place of $\epsilon$, under the square root. Of course
this coefficient becomes $2$, viewed from the place of
$\kappa_{1}$, or even from the place of $\Lambda$, that is to say
outside the square root. The interpretation of this factor $4$
depends on how the quantum equation is established in the context
of unification, and depends on the origin of the dependence of the
gravitational coupling $G$ on $\epsilon$. Furthermore, the problem
of these coefficients would not have appeared in other quantum
cosmological models, the relation (3.5) does not make appear any
square root for example. In fact, equation (1.5) renders difficult
to compare general relativity and our quantum equation. To simply
the problem, we just recall that the net effect of $\Lambda$ is to
multiply the right hand side by $2$. This multiply $\kappa_{1}$ by
$2$ or $\epsilon$ by $4$. In general relativity, multiplying the
right hand side by $2$ would have multiply $\epsilon$ by $2$, and
this behavior, inserted in the quantum (3.16), would have multiply
$\kappa_{1}$ by $\sqrt{2}$. So, we have a incertitude on the value
of $\kappa_{1}$ of a factor like $\sqrt{2}$ (or $2$ because
$\Lambda$ also multiply $\kappa_{1}$ by $2$). Maybe equation (3.5)
can solve this problem, however simple relations below will show
that the fundamental constant is $\kappa_{0}$, given by the
relation :
\begin{equation}\kappa_{0}=2\sqrt{2}\kappa_{1}\end{equation}
\subsubsection{Value of the kappas}
The flatness problem is equivalent to prove that the total energy
density of matter, $\Omega=\Omega_{TOT}$, that is to say the
energy density when dark energy is taken into account, has a
present value near unity. In classical gravity, the value of
$\Omega_{TOT}$ takes the form
\begin{equation}\Omega=\Omega_{TOT}=\frac{8\pi G\epsilon_{app}}{3H^{2}}\end{equation}
A reasonable relation between $G$ and $\kappa_{1}$ is found by
comparing general relativity and quantum gravity. In quantum
gravity, we have :
\begin{equation}\frac{K}{a^{2}}+H^{2}=\frac{2}{3}\kappa_{1}\sqrt{\epsilon}\end{equation}
In general relativity the relation was :
\begin{equation}\frac{K}{a^{2}}+H^{2}=\frac{8\pi
G}{3}\epsilon_{app}\end{equation} We also have
$\epsilon_{app}=4\epsilon$, which is a consequence of the quantum
equation of gravity. Inserting this relation in (3.28) (this
relation is also valid in the classical case in virtue of the
observations), we obtain :
\begin{equation}\frac{K}{a^{2}}+H^{2}=\frac{32\pi G}{3}\epsilon\end{equation}
Comparing (3.27) and (3.29) we find :
\begin{equation}\kappa_{1}=16\pi G\sqrt{\epsilon}\end{equation}
Using (3.23), we find : \begin{equation}\kappa_{1}=\frac{3\Omega
H^{2}}{2\sqrt{\epsilon}}=16\pi G\sqrt{\epsilon}\end{equation} and
$\kappa_{1}$ is also the square root of the two expressions in
(3.31) which leads to :
\begin{equation}\kappa_{1}=\sqrt{24\pi}G^{1/2}(\Omega^{1/2}H)\end{equation}
Using finally (3.25) we obtain the value of $\kappa_{0}$ :
\begin{equation}\kappa_{0}=\sqrt{192\pi}G^{1/2}(\Omega^{1/2}H)\end{equation}
Or, we can deduce the value of Newton's constant :
\begin{equation}G=\frac{1}{192\pi}\frac{\kappa_{0}^{2}}{(\Omega^{1/2}H)^{2}}\end{equation}
Using the observed values of $\Omega_{TOT}$ and $H$ (Bennett,
2003, $[5]$), we find :
\begin{equation}\kappa_{1}=1.087\times10^{-60}\end{equation}
and :
\begin{equation}\kappa_{0}=3.074\times10^{-60}\end{equation}
\subsection{The flatness problem}
We have to prove that the expression for $\Omega_{TOT}$ tends to a
finite value when $t\rightarrow+\infty$. We know that the present
value of $\dot{a}$ is $\dot{a}=\lambda$. We thus find
$$\frac{1}{a^{2}}=\frac{H^{2}}{\lambda^{2}}$$ Finally, defining
\begin{equation}\mu=\frac{\lambda^{2}}{\lambda^{2}+1}\end{equation}
we obtain :
$$\frac{1}{\mu}H^{2}=\left(\frac{K}{\lambda^{2}}+1\right)H^{2}=\left(\frac{K}{\dot{a}^{2}}+1\right)H^{2}=\frac{K}{a^{2}}+H^{2}=\frac{2}{3}\kappa_{1}\sqrt{\epsilon}$$
such that :
\begin{equation}\frac{1}{\mu}H^{2}=\frac{2}{3}\kappa_{0}\sqrt{\epsilon}\end{equation}
We now use the value of $\Omega_{TOT}$ in (3.23) (which also can
be deduced from (3.26) and (3.30)), and find :
\begin{equation}\Omega=\Omega_{TOT}=\frac{2\kappa_{1}\sqrt{\epsilon}}{3H^{2}}=\frac{1}{\mu}\geq1\end{equation}
Here we recall that we are necessarily in the closed model
(section 3.4). We thus have :
\begin{equation}\frac{1}{\mu}=\frac{1}{\lambda^{2}}+1\end{equation}
in such a way that $1/\mu\approx1$ and $1/\mu\geq1$. When
$\lambda$ is not used anymore to note the present value of
$\dot{a}$, but rather its limit when $t\rightarrow+\infty$, our
result is not the present value of $\Omega$ but its limit value.
The present value have been observed to be, in the context of the
standard cosmological model (Bennett and al., 2003) :
\begin{equation}\Omega=1.02\pm0.02\end{equation}
To find $\Omega=1.02$ in the quantum context, we need the present
value of $\dot{a}$ to be
$$\dot{a}_{pr}=\lambda=7.07$$ and to find the greatest possibility
$\Omega=1.04$ we need
$$\dot{a}_{pr}=\lambda=5$$
A remark can be made : if it can be observed, in our universe,
distances of the order of $200Mpc$, and if the $cH^{-1}$ distance
is about $4000Mpc$, we then are sure that $\dot{a}\geq1/20=0.05$.
That the universe could be one hundred times bigger than this
minimum value does not seem a priori to be ruled out by any
experiment, and only very small values of $\dot{a}$ are ruled out.
We emphasize that no theory before the one presented here has been
able to explain so precisely the value $\Omega=1.02$ coming from
observations.
\section{Proof of a conjecture of De Broglie}
\subsection{Probing the structure of the particle using the conjecture}
We suppose here De Broglie's famous hypothesis (De Broglie, 1963,
1963, $[6]$), that inside a fundamental particle, the temperature
is equal, or at least proportional to the temperature. We further
suppose that the total number of particles stays almost constant
in the universe, and we prove that we necessarily are in one of
our quantum models. If the former hypotheses are strictly
verified, we even are in the $\theta\rightarrow+\infty$ model.
Indeed, to prove this, we place ourselves in the case in which
matter, we mean the set made of the usual fundamental particles we
know, is non relativistic. We emphasize that these particles are
the particles we know, and are supposed to be almost pointlike.
Probing their structure means determining in they are made of
relativistic stuff or not. They can be made for example of
confined photons, in which case they still can be non relativistic
in the sense that their global relative speeds are non
relativistic, but inside, the value of the pressure still can be
$p=\epsilon/3$. We see that the occupation number of this non
relativistic gas, constituted of (almost) pointlike particles, is
necessarily
\begin{equation}N_{m}\sim\exp[-p^{2}/2mT_{m}]\end{equation}
where we take $k=1$ for the Boltzmann constant, where $T_{m}$
denotes the temperature of matter, and $p$ are the particle
momenta. As proved in Peebles, 1993, $[7]$, the particle momenta
are proportional to $1/a$, so we find that the occupation number
is constant on the condition :
\begin{equation}mT_{m}\sim\frac{1}{a^{2}}\end{equation}
We now probe the structure of particles, and use that the mass
should be proportional to the temperature. With this further
relation, $T_{m}\sim m$, We find from (4.2) :
\begin{equation}m\sim T_{m}\sim\frac{1}{a}\end{equation}
This gives a behavior of the mass, analog to what we found from
our quantum equation of gravity. We deduce even more, because this
specific case in which $m\sim1/a$, corresponds to
$\epsilon\sim1/a^{4}$, since the mass is equal to the energy for
the non relativistic gas. Indeed, we have
$$\epsilon=\frac{N_{m}m}{V}\sim\frac{N_{m}m}{a^{3}}\sim\frac{1}{a^{4}}$$ where $V$ is
the total volume of the universe, because of (4.3), and also
because $N_{m}$ is constant. Using the conservation of entropy,
this relation for $\epsilon$ means $p=\epsilon/3$, where now $p$
is the pressure, and not anymore the momenta of the particles, as
in (4.1). This last relation means that we are in the relativistic
case. This does not contradict our first hypothesis, that the
particles which we considered were non relativistic, but proves
that their structure is relativistic. For example, we can suppose
that these particles are made of more fundamental, relativistic
and confined constituents, or simply made of photons. We also can
leave this structure undefined, even if still relativistic, for
the time being. From this relation $p=\epsilon/3$, and from
relation (3.3) :
\begin{equation}p=\frac{\epsilon}{3}-\frac{\sqrt{\epsilon}}{2\kappa_{1}\theta}\end{equation}
proved in $[1]$, we deduce that we are in the case
$\theta\rightarrow+\infty$, which unifies all quantum models in a
limit case.
\subsubsection{Proof of De Broglie's conjecture}
Inversely, starting from the hypothesis that
$\theta\rightarrow+\infty$, using equation (4.4), we deduce that
$p=\epsilon/3$, and thus that :
\begin{equation}\epsilon\sim1/a^{4}\end{equation}
using the conservation of entropy. The particles we know are not
observed to possess these relative relativistic velocities, which
means that this relativistic behavior is not the consequence of
the relative velocities of the usual fundamental particles, but
rather a consequence of their structure. We thus can adopt the
model which imposes that these known particles are made of more
fundamental relativistic ones, the best candidate being of course
the photon (we recall that the graviton will soon be expected to
have no existence). Depending on the cosmological model, the
relation $p=\epsilon/3$ is only approximate, or on the contrary,
in the case $\theta\rightarrow+\infty$, rigorously verified. Since
these baryons can be approximated by the picture of a non
relativistic gas, we obtain (4.1), and the additional condition of
conservation of their total number gives (4.2), where we still use
$[7]$. Since the total mass $N_{m}m$ is proportional to the total
energy $\epsilon a^{3}$, and since $N_{m}$ is constant, we deduce
from (4.5) the following relation, which is one part of (4.3) :
\begin{equation}m\sim\frac{1}{a}\end{equation}
We now combine (4.2) and (4.6), and obtain $T_{m}\sim1/a$. The
last relation, compared to (4.6), gives :
\begin{equation}T_{m}\sim m\end{equation}
Finally, the equation of quantum gravity gives a proof of the
relation between the masses and the temperature, that is to say a
proof of De Broglie's conjecture.
\subsection{Conclusion concerning the masses}
As proved in $[1]$, the quantum equation of gravity governs at the
same time the large scale of the universe and the structure of
particles. We thus look for relations between the masses of the
particles and the cosmological parameters. $H$ has the dimension
of an energy, a classical mass $m$ have the dimension of an energy
too, and we just saw that the associated quantum mass, noted
$\tilde{m}$ from now on, is dimensionless. We thus look for a
relation like :
\begin{equation}m=\tilde{m}H\end{equation} or like :
\begin{equation}m=\tilde{m}(\Omega^{1/2}H)\end{equation}
where $\tilde{m}$ is the true constant of nature, in particular
the true value of the mass, constant in  respect to the
cosmological time. The former relations are identical in the case
$\Omega=1$, and are very near from each other in the case
$\Omega\approx1$. Relation (4.9) corresponds to De Broglie's
statement $m\sim T$. For the time being, we look at every
possibility, and each of these possibilities corresponds to a
different cosmological model, as we prove below. From (3.23), we
deduce :
\begin{equation}\Omega
H^{2}=\frac{2}{3}\kappa_{1}\sqrt{\epsilon}\approx\kappa_{1}\sqrt{\epsilon}\end{equation}
We also have
\begin{equation}G\sim\frac{\kappa_{1}}{\sqrt{\epsilon}}\end{equation}
up to a constant $16\pi$. So we see that
\begin{equation}G^{1/2}\approx\kappa_{1}(\Omega^{1/2}H)^{-1}\end{equation}
The relations (4.9) and (4.12) permit us to retrieve that the
gravitational charges are constant in time, taking into account
that the true gravitational charges are the combinations
$G^{1/2}m$ and not $G$ or $m$. In this case, the model does not
predict any observable variation of the gravitational charges. So
this model does not predict any variation of the intensity of the
gravitational interaction. In order to investigate all quantum
models, the other formula (4.8) would, with this behavior of $G$,
be a model with small variations of the intensity of the
gravitational interaction. However, we still can use (4.8) and
apply at the same time the principle of constancy of the intensity
of the gravitational interaction. In this case, to obtain the
constancy of the strength of the true gravitational charge
$G^{1/2}m$, we have to suppose that $G$ is strictly proportional
to $H^{-2}$, and we find :
\begin{equation}G\sim\frac{\kappa_{1}^{2}}{H^{2}}\end{equation}
At the end, we already have four different cosmological models,
two choices for the behavior of $G$, and two choices for the
behavior of $m$. The former relations shall be proved several
times in this work, up to a term containing only $\dot{a}$, and we
saw in $[1]$ that in the quantum models, $\dot{a}$ has a present
value around unity. We notice that, up to such a term, (4.11) and
(4.13) (respectively (4.8) and (4.9)) are the same. In the limit
case, when $\theta\rightarrow+\infty$, we find that $\dot{a}$ is
constant, and all models tend to the same limit. This behavior of
the masses, (4.8) or (4.9) or any other formula corresponding to
(4.8) up to a factor containing only $\dot{a}$, corresponds to a
solution to the mass gap problem of the Clay Mathematics
Institute. If the masses are proportional to $H$, and if $H$ tends
to zero when $t\rightarrow+\infty$, we prove that the answer to
the mass gap problem is no, that there can be no proof of a mass
gap from the side of gauge theories, since there is no physical
mass gap at all. On the contrary, if we try to apply the mass gap
not anymore to $m$ but to $\tilde{m}$, there is a mass gap, as we
will prove, using only the Heisenberg's incertitude relations.
Furthermore, the term containing only $\dot{a}$ up to which the
formula is proved, will not change enough the behavior of the
masses to be able to invalidate this proof.
\section{A new constant of physics}
\subsection{Interpretation of Hubble's constant}
In order to interpret the constant of gravitation, and later
gravitation itself, we first need to interpret Hubble's constant,
and this can be done remembering Heisenberg's uncertainty
relations. From the relation
\begin{equation}\Delta E \Delta t\geq1\end{equation}
we see that the uncertainty, in the observed value of any energy
$E$, is linked to the interval of time of the observation $\Delta
t$. But of course, this interval of time of the observation cannot
be greater than the age of the existing universe itself. In the
standard cosmological model, this value of the age of the universe
is about $H^{-1}$. So we finally obtain :
\begin{equation}\Delta E\geq\frac{1}{t}\approx H\end{equation}
Landau, in $[8]$, explains that such an Heisenberg uncertainty
relation proves that any strictly positive energy $E$ cannot be
smaller that $H$. Indeed, any energy below this value could not be
distinguished from zero, because of the uncertainty. We conclude
that $H$ is the smallest energy, strictly positive, ever possible
in the universe. This, of course, corresponds to an energy gap,
coming from cosmological necessities.
\subsection{Interpretation of the constant of gravitation} We can
give an interpretation of the constants $\kappa_{0}$ and
$\kappa_{1}$ (we stick here to $\kappa_{0}$), analyzing (3.33) :
\begin{equation}\kappa_{0}\approx\sqrt{192\pi}\frac{H}{M_{Pcl}}\end{equation}
Here, $M_{Pcl}=G^{-1/2}$ is the classical Planck mass, and because
$\Omega^{1/2}\approx1$, we leave this term for the moment. To
eliminate the term $192\pi$, we note the quantum Planck mass :
\begin{equation}M_{P0}=\frac{1}{\sqrt{192\pi}}M_{Pcl}\end{equation}
and we obtain :
\begin{equation}\kappa_{0}=\frac{H}{M_{P0}}\end{equation}
To analyze (5.5), we notice that $M_{Pcl}$ has always been
considered as an energy scale beyond which quantum gravity will
come into play. $(M_{Pcl})^{-1}$, written as a distance, is
believed to be the distance at which space-time breaks down. Now,
in the former formula, there is $H$ in the numerator, which is the
smallest energy possible in the universe. On the other hand,
$\kappa_{0}$, and its powers with exponents between $-1$ and $1$
(see below), controls ratios of intensities of fundamental
objects, for example fundamental interactions. The interpretation
of the denominator is now clearer : this is the greatest energy
possible of one particle, that is the physical cut-off in Feynman
graphs which has been needed for so long, and $\kappa_{0}$ is the
tiniest ratio possible, the smallest energy possible of one
particle divided by the greatest energy possible of one particle.
For this reason, $H$ plays also the role of the infrared cut-off
in Feynman integrals. $\kappa_{0}$ is the physical value
corresponding to the breakdown that the mathematical real axis
itself should have when it is applied to describe physics, and we
believe that $\kappa_{0}$ should be used to try to explain the
"quantum fact".
\section{Electromagnetism and Gravitation}
\subsection{Introduction}
In the large number hypothesis, we can find a relation between the
mass of the pion and a cosmological parameter like Hubble's
constant. At this stage, this relation could still be considered
to be a mere coincidence, or on the contrary, to have a true
physical meaning. We now prove that, not only this relation has a
true physical meaning, but also that it belongs to a vast web of
relations allowing us to compute all parameters of the standard
model of particle physics. The relations of this web cannot be
coincidences, first because of their great number, and second
because of their astonishing accuracy.
\subsection{Renormalization theory and tensorial quantum gravity}
The principle of construction of the tensorial equation of gravity
was that Newton's gravitational constant should depend on energy,
to behave exactly like the couplings of the other interactions,
after they have been renormalized. In the equation of quantum
gravity, Newton's constant is proportional to :
\begin{equation}G\sim\kappa(\epsilon)\sim\frac{\kappa_{0}}{\sqrt{\epsilon}}\end{equation}
The typical running of a coupling constant after renormalization,
for example the running of $\alpha_{em}$ in Quantum
Electrodynamics, is the following. The variations of $\alpha$,
between two energies $\mu_{1}$ and $\mu_{2}$, are described by :
\begin{equation}\frac{1}{\alpha(\mu_{1})}=\frac{1}{\alpha(\mu_{2})}-\frac{2}{3\pi}\ln\left(\frac{\mu_{1}}{\mu_{2}}\right)\end{equation}
Clearly, it appears that the behaviors in (6.1) and in (6.2) are
quite different. The first idea was to obtain a running of $G$
analog to the running of $\alpha$. We will see just below how this
problem will find its solution.
\subsection{The relation}
The experimental value of the coupling of electromagnetism
$\alpha_{em}$ is given by
$$\frac{1}{\alpha_{em}}=137.036$$ with great accuracy. Here,
$\alpha_{em}$, which depends on energy, is observed at the energy
of the mass of the electron. Now, using $\Omega=1.02$ and $h=0.71$
for the value of Hubble's constant (we recall $h$ is defined by
the formula $H=100.h.km.s^{-1}.Mpc^{-1}$), we calculated the
coupling constant $\kappa_{0}$ and we found in (3.35) and (3.36) :
$\kappa_{0}=2\sqrt{2}\kappa_{1}=2\sqrt{2}\times1.087\times10^{-60}$.
Finally :
\begin{equation}\kappa_{0}=3.074\times10^{-60}\end{equation}
To find what to do, we remember the interpretation we gave for
$\kappa_{0}$. We obtained the relation
$$\kappa_{0}=\frac{H}{M_{P0}}$$
We interpreted $H$ as the smallest strictly positive energy that
can carry one particle, and $M_{P0}$, the quantum Planck mass
related to $\kappa_{0}$, as the greatest energy that can be
carried by one particle, that is to say the true physical cut off
in Feynman integrals. $\kappa_{0}$ was the dimensionless ratio
representing the smallest physical ratio ever possible in our
universe. We have found this way a solution to the problem of
renormalization, since we have now a true physical cut off in
Feynman integrals. Especially, in (6.2), there is no reason
anymore to suppose that running $\alpha$ diverges when, for
example, $\mu_{1}\rightarrow+\infty$, simply because $\mu_{1}$ can
never be greater than the physical cut off. In other words, in
(6.2), the ratio $\mu_{1}/\mu_{2}$ can never exceed the greatest
ratio ever possible in the universe, which is, according to our
interpretation, the ratio of the quantum Planck mass to the
smallest energy defined by Hubble's constant, since $H$ is the
infrared cut-off in Feynman integrals. In other words, this
greatest ratio is $(\kappa_{0})^{-1}$. If our views are correct,
the logarithm of this ratio should be of the order of the two
other terms in equation (6.2), that is to say around $137$. This
way, we would have proved that our physical cut off has exactly
the right order of magnitude to make quantum corrections of the
numerical order of the quantities they correct. Furthermore, if
the relation between $(\kappa_{0})^{-1}$ and $1/\alpha_{em}$ is
made via a logarithm, this explains perfectly the differences of
behaviors in (6.1) and (6.2). This would explain why the running
couplings possess logarithms in the quantum theory : the running
of $G$ on energy is a power law, this power law explains why we
did not at first find a dimensionless Newton's constant, because
this power law breaks the property of being dimensionless.
Furthermore, taking the logarithm, this power law transforms into
the usual running of the couplings after renormalization. At this
stage, we take the logarithm of
\begin{equation}\kappa_{0}=3.074\times10^{-60}\end{equation}
and find :
$$\ln\left(\kappa_{0}^{-1}\right)=137.0321$$
This represents a relative uncertainty of only $2.8\times10^{-5}$
from the observed value of $\alpha_{em}$, so we have found the
relation between the coupling constants of electromagnetism and
gravitation :
\begin{equation}\ln\left(\kappa_{0}^{-1}\right)=\frac{1}{\alpha_{em}}\end{equation}
This proves that the constant $\kappa_{0}$ does not only contain
the data of the coupling constant $\alpha_{em}$, but also the data
of the mass of the electron, since at another energy,
$\alpha_{em}$ takes a different value. This explains the large
number hypothesis, because from the value $1/\alpha_{em}=137.036$,
we find :
\begin{equation}\kappa_{0}=\exp[-1/\alpha_{em}]\approx3\times10^{-60}\end{equation}
In particular the term $10^{20}$ which appeared everywhere in this
hypothesis was nothing else that $\exp[-1/3\alpha_{em}]$.
\section{The mass of the electron}
From the last relation, we can calculate the value of the
constants of gravitation $\kappa_{0}$ and $\kappa_{1}$. We have of
course
\begin{equation}\kappa_{0}=\exp\left[-\frac{1}{\alpha_{em}}\right]\end{equation}
and using the value of $\alpha_{em}$ given by CODATA $[9]$ :
\begin{equation}\frac{1}{\alpha_{em}}=137.035999679(94)\end{equation} we find
\begin{equation}\kappa_{0}=3.06211514(29)\times10^{-60}\end{equation}
Using $\kappa_{1}=\kappa_{0}/2\sqrt{2}$, we obtain :
$$\kappa_{1}=1.08262119(11)\times10^{-60}$$
We notice that from a relative uncertainty on the constant
$\alpha_{em}$ which was $6.8\times10^{-10}$, the consequence of
the exponential law is that the relative uncertainty on the
gravitational coupling is $9.4\times10^{-8}$.
\subsection{The value of Hubble's constant}
From the former relation, and the particular model we used in
$[1]$, we are able to predict the value of Hubble's constant with
great accuracy. More precisely, we can compute the expression
$\Omega^{1/2}H$ to a relative precision of $1\times10^{-4}$, and
$H$ to a precision of $10^{-2}$, while the relative precision
given by Bennett and al., 2003, was $5\times10^{-2}$. We recall
here that these formulas are available only in the particular
quantum  model we consider, and that many other models are
possible, as we already showed. However, this kind of calculation
shall permit us to determine which model we should choose,
comparing the predictions with experiment. Here we use (3.33)
$\kappa_{0}=\sqrt{192\pi}(\Omega^{1/2}H)\sqrt{G}$ and find :
\begin{equation}\Omega^{1/2}H=\frac{G^{-\frac{1}{2}}}{8\sqrt{3\pi}}\exp\left[-\frac{1}{\alpha_{em}}\right]
=\frac{M_{Pcl}}{\sqrt{192\pi}}\exp\left[-\frac{1}{\alpha_{em}}\right]\end{equation}
From the list of the CODATA $[9]$ recommended values of the
fundamental constants, we obtain :
$$M_{Pcl}=1.220892(61)\times10^{19}GeV$$
We compute :
\begin{equation}\Omega^{1/2}H=1.522205(76)\times10^{-42}GeV\end{equation}
To obtain the value of $H$, we recall that the value of $\Omega$
is between $1$ and $1.04$. We find for $H$ :
\begin{equation}H=1.507(15)\times10^{-42}GeV\end{equation}
We find the value of $h$ :
\begin{equation}h=0.707(7)\end{equation}
which is again almost at the center of the interval of
uncertainties of the cosmological observations (Bennett and al.,
2003). We emphasize once again that these predictions only test
one of our numerous quantum equations of gravity. Another problem
is that the observed value of $\Omega$ may be model dependent.
However, we stick to the observed interval of values since we
showed in $[1]$ that the quantum model gives theoretical
predictions for the value of $\Omega$ which lie exactly in the
same interval.
\subsection{A formula for the mass of the pion}
We know there exists an approximate relation for the mass of the
pion :
\begin{equation}m_{\pi}\approx\left(\frac{\bar{h}^{2}H}{Gc}\right)^{\frac{1}{3}}\end{equation}
We now use now what we know, that is to say that the true mass of
the pion is $\tilde{m}_{\pi}$, in the relation
$m_{\pi}=\tilde{m}_{\pi}H$. We also use relation (3.34)
$$G=\frac{1}{192\pi}\frac{\kappa_{0}^{2}}{H^{2}}$$
(here we neglect the fact that $\Omega$ is not strictly equal to
$1$, which is an error less than only $2$ percents). Replacing in
(7.8), we obtain :
\begin{equation}m_{\pi}=\left(\frac{H}{G}\right)^{1/3}\approx
\left(\frac{(\Omega^{1/2}H)}{G}\right)^{1/3}=(192\pi)^{1/3}\left(\kappa_{0}\right)^{-\frac{2}{3}}(\Omega^{1/2}H)
\approx\left(\kappa_{0}\right)^{-\frac{2}{3}}(\Omega^{1/2}H)\end{equation}
We recall that the term $\Omega^{1/2}$ can be present or absent in
the equation, depending on the quantum model we choose. Thus, the
simple relation which remains is :
\begin{equation}\tilde{m}_{\pi}\approx(\kappa_{0})^{-2/3}\end{equation}
We left in the last formula the term $(192\pi)^{1/3}\approx8.45$
because first it is near unity, and second because it seems to be
a consequence of the formula giving the classical Newton's
constant $G$. So it appears that this term still belongs to (semi)
classical physics.  Without the numerical factor $192\pi$, but
with $\Omega$, we find for the mass of the pion the value
$7.22MeV$. This is in the domain of the masses of the known
particles, but now much nearer to the mass of the electron. With
the coefficient $192\pi$ we find the value $61MeV$, which is half
the value of the mass of the pion. This confirms that the powers
of $\kappa_{0}$ determine the fundamental quantities in the
universe.
\subsubsection{The tininess of gravitation}
We now show that the former value for the masses explains the
ratio of the gravitational to the electromagnetical forces. We
call $\beta_{e}$ and $\beta_{p}$ the values of $\beta$
corresponding respectively to the masses of the electron and the
proton, in the formula : $$\tilde{m}=(\kappa_{0}^{-1})^{\beta}$$
From the former section, we can say in first approximation that we
have : $\beta_{e}=\beta_{p}=2/3$. We compute :
$$\kappa_{0}^{2/3}=2.1\times10^{-40}$$ We know from (3.34) that :
$$G=\frac{1}{192\pi}\frac{\kappa_{0}^{2}}{H^{2}}$$ and that
$$e^{2}=4\pi\alpha_{em}$$
So the ratio
\begin{equation}\frac{Gm_{p}m_{e}}{e^{2}}
=\frac{{\kappa_{0}}^{2}({\kappa_{0}}^{-1})^{\beta_{p}}({\kappa_{0}}^{-1})^{\beta_{e}}}{768\pi^{2}\alpha_{em}}\approx{\kappa_{0}}^{2/3}\end{equation}
Indeed, we neglect the denominator, and we use
$\beta_{p}=\beta_{e}=2/3$. We arrive at :
$$\frac{Gm_{p}m_{e}}{e^{2}}\approx{\kappa_{0}}^{2/3}\approx2\times10^{-40}$$
Inversely, we could have started from the last relation, and using
(7.11) plus the condition $m_{p}\approx m_{e}$, that is to say
$\beta_{p}=\beta_{e}$, we would have obtained :
$$2-\beta_{e}-\beta_{p}=2-2\beta_{e}=\frac{2}{3}$$ and thus :
\begin{equation}m_{e}\approx m_{p}\approx{\kappa_{0}}^{-2/3}H\end{equation}
which is the right computation of the masses. Once again, a power
of the constant $\kappa_{0}$ gives the ratio between the
intensities of two interactions, and once again $\kappa_{0}$
appears as the principle of unification of the interactions. The
formula for the masses and the formula for the ratio of the two
interactions comes from only one principle, and are equivalent, as
variations about different ways of writing $2/3$ :
$2-2/3-2/3=2/3$.
\section{The mass of the electron}
If we look carefully at our relation :
$$\frac{Gm_{e}m_{p}}{\alpha_{em}}={\kappa_{0}}^{\frac{2}{3}}$$
We note that $$G=\frac{1}{192\pi}\frac{{\kappa_{0}}^{2}}{H^{2}}$$
and notice that $192\pi\approx603$, which, up to a factor $3$, is
almost the famous ratio $1836$ of the mass of the proton to the
mass of the electron. Forgetting this factor $3$, we write
$m_{p}\approx192\pi m_{e}$, and replacing in the equation we find
\begin{equation}\left(\frac{m_{e}}{H}\right)^{2}=\alpha_{em}{\kappa_{0}}^{-\frac{4}{3}}\end{equation}
which gives
\begin{equation}m_{e}=\sqrt{\alpha_{em}}{\kappa_{0}}^{-\frac{2}{3}}H\end{equation}
We know that we can also write :
\begin{equation}m_{e}=\sqrt{\alpha_{em}}{\kappa_{0}}^{-\frac{2}{3}}(\Omega^{1/2}H)\end{equation}
Using (7.2), (7.3), (7.5) and (8.3), we find $m=0.617MeV$, which
is quite near the observed mass of the electron. We have to remind
ourselves that there can be also a coefficient $\sqrt{2}$
appearing in the formula. More precisely, if we look at the last
relation, we find that it is the simplest relation possible for
the mass of the electron. In the context of unification, the
formula for the mass should be given only via the ratio $m/e$ of
the mass of the electron to its electromagnetic charge, and not
directly via $m$ alone. So the simplest formula, for the mass of
the electron, is
\begin{equation}\frac{\tilde{m}_{e}}{\sqrt{\alpha_{em}}}=(\kappa_{0})^{-2/3}\end{equation}
where $\sqrt{\alpha_{em}}$ plays naturally the role of the charge
of the electron. This is exactly formula (8.3), which gives a
value just a little too high. However, the complete gravitational
charge is clearly proportional to
$\kappa_{0}\tilde{m}_{e}=\sqrt{\alpha_{em}}(\kappa_{0})^{1/3}$. In
the quantum equation of gravity the effective coupling was
$\kappa_{1}$, and not $\kappa_{0}=2\sqrt{2}\kappa_{1}$, so the
true gravitational charge should not be
$\sqrt{\alpha_{em}}(\kappa_{0})^{1/3}$ but instead
$\sqrt{\alpha_{em}}(\kappa_{1})^{1/3}$, which is :
$$\frac{\sqrt{\alpha_{em}}}{\sqrt{2}}(\kappa_{0})^{1/3}$$
Thus, the true formula for the mass of the electron is :
\begin{equation}m_{e}=\sqrt{\frac{\alpha_{em}}{2}}{\kappa_{0}}^{-\frac{2}{3}}(\Omega^{1/2}H)\end{equation}
and this way, we find the value $m_{e}=0.4360401MeV$. We add to
this value the electron self-energy mass shift in second-order QED
perturbation theory (Mandl, Shaw, $[10]$). We find :
$$\delta
m=\frac{3e^{2}m}{8{\pi}^{2}}\ln\left(\frac{\Lambda}{m_{e}}\right)$$
We take the mass $\Lambda$ to be determined by
$\beta_{\Lambda}=1$, which is the additional hypothesis we made
about the physical cut-off in Feynman integrals. We find then
\begin{equation}\delta
m=\frac{3m}{2\pi}(\beta_{\Lambda}-\beta_{e})\end{equation} Indeed,
the classical formula is
\begin{equation}\delta m=\frac{3m}{2\pi}\alpha_{em}\ln\frac{\Lambda}{m_{e}}\end{equation}
and
$$\Lambda=(\kappa_{0}^{-1})^{\beta_{\Lambda}}(\Omega^{1/2}H)$$
also $$m_{e}=(\kappa_{0}^{-1})^{\beta_{e}}(\Omega^{1/2}H)$$ and
$$\alpha_{em}\ln[(\kappa_{0})^{-1}]=1$$ from the formula of
unification. To compute the value of $\beta_{e}$, from
$m_{e}=({\kappa_{0}}^{-1})^{\beta_{e}}(\Omega^{1/2}H)$ we have to
notice that the value of $m_{e}$ that has really been computed is
$0.436MeV$ and not the observed $0.511MeV$. So we have to use this
$0.436$ in the former formula to obtain the true value of
$\beta_{e}$. This way, the new renormalized mass of the electron
is
\begin{equation}m_{e}=0.5097MeV\end{equation} which, compared to the observed
\begin{equation}m_{e}=0.511MeV\end{equation} is inside the interval of uncertainties, taking
into consideration that we have left aside corrections to higher
orders. The relative uncertainty is $2.6\times10^{-3}$, which is
very good.
\section{Weak, Strong and Gravity}
\subsection{The Fermi constant}
The fermi constant $G_{F}$ has a simple expression in terms of
$\kappa_{0}$. First we know that Newton's constant $G$ reads :
\begin{equation}G=\frac{1}{192\pi}\frac{\kappa_{0}^{2}}{(\Omega
H^{2})}\end{equation} Second we know, from the CODATA $[9]$
recommended values, that :
\begin{equation}G_{F}=1.16637(1)\times10^{-5}(GeV)^{-2}\end{equation}
To obtain a dimensionless Fermi constant, which we still note
$G_{F}$, we have to multiply this expression by $\Omega
H^{2}=\left(1.522205(76)\times10^{-42}GeV\right)^{2}$. The
dimensionless Fermi constant is equal to :
\begin{equation}G_{F}=2.70260533\times10^{-89}\end{equation}
In this value, appears clearly $(\kappa_{0})^{3/2}$, and we find :
\begin{equation}G_{F}=5.04371(50)(\kappa_{0})^{3/2}=\tilde{G}_{F}(\kappa_{0})^{3/2}\end{equation}
Once again, up to a factor of order unity, one power of
$\kappa_{0}$ determines another fundamental constant of nature.
$\kappa_{0}$ governs the intensity of electromagnetism by its
logarithm, giving $\alpha_{em}$, it governed the strong
interactions via the mass of the pion by $(\kappa_{0})^{-2/3}$,
and now governs the Fermi constant via $\kappa_{0}^{3/2}$. The
value $\tilde{G}_{F}=5.04371$ should be fundamental, and
expressible in a simple manner. Also, when the dimensionless Fermi
constant is not written with $\kappa_{0}$ but with $\kappa_{1}$,
we find :
$$G_{F}=3\times0.99967(10)\times(\sqrt{2}\kappa_{0})^{3/2}$$
\begin{equation}=24\times0.99967(10)\times(\kappa_{1})^{3/2}\end{equation}
It is astonishing to find in this expression a number so near
unity, even if the value $1$ is ruled out by the precision of
experiments. However, it appears that the powers of $\kappa_{0}$
clearly govern all values of the standard model of particles
physics, which are now calculable up to a constant near unity, and
that we need another theory at this stage to compute this
constant. The same thing is going to happen in the calculation of
the masses of the $W$ and $Z$ particles.
\subsection{The masses of the $W$ and $Z$ particles}
The theory of weak interactions exhibits the formula :
\begin{equation}\frac{G_{F}}{\sqrt{2}}=\left(\frac{g_{W}}{m_{W}}\right)^{2}\end{equation}
where $g_{W}$ is the coupling constant of the weak interactions,
and $m_{W}$ is the mass of the $W$ particle. So we obtain :
\begin{equation}g_{W}=1.58803(5)m_{W}(\kappa_{0})^{3/4}\end{equation}
From (9.6), the $W$ particle verifies the relation :
\begin{equation}m_{W}=g_{W}\left(\frac{G_{F}}{\sqrt{2}}\right)^{-\frac{1}{2}}\end{equation}
and since the only power of $\kappa_{0}$ appearing in the last
formula is $(\kappa_{0})^{3/2}$, coming from $G_{F}$, we find that
$m_{W}$ is governed by the power : $(\kappa_{0})^{-3/4}$. So is
$m_{Z}$, since the two particles have about the same mass. The
simplest formula, that should give $m_{W}$, as the power
$\kappa_{0}^{-3/4}$, is :
\begin{equation}m_{W}=\frac{\sqrt{\alpha_{em}}}{2^{3/8}}(\kappa_{0})^{-3/4}(\Omega^{1/2}H)\end{equation}
The complete gravitational charge is
$\kappa_{0}m_{W}\sim\kappa_{0}^{1/4}=2^{3/8}\kappa_{1}^{1/4}$,
which explains the factor $2^{3/8}$ in the denominator. In this
formula, the term $\alpha_{em}$ is linked to the electric charge
of the $W$ particle, and we have to take its value at the energy
$m_{W}$. The theory of unification of electro-weak interactions
gives us relations between the masses of the $W$ and the $Z$
particles. In Pich, 1995, $[11]$, the mass shifts of these two
particles are computed. As far as the most important QED and QCD
corrections to the mass are concerned, these corrections affect
the relation between $m_{W}$ and $m_{Z}$, but we do not know if in
a more unified theory these changes would affect more specifically
$m_{W}$ or $m_{Z}$. In $[11]$, $m_{Z}$, especially, is taken from
experiment and kept constant all along the calculation. We do the
same here, in other words $m_{Z}$, is not affected by these
quantum corrections. The observed values of the two masses are
$m_{W}=80.403GeV$ and $m_{Z}=91.1887GeV$ (Yao and al. 2006,
$[9]$). We apply formula (9.9) to $m_{Z}$, an allow for a
coefficient $\lambda_{Z}$, yet unknown and to be determined. We
obtain :
\begin{equation}m_{Z}=\lambda_{Z}\frac{\sqrt{\alpha_{em}}}{2^{3/8}}(\kappa_{0})^{-3/4}(\Omega^{1/2}H)\end{equation}
To compute this theoretical value of the mass, we have to take the
value of $\alpha_{em}$ at the energy $m_{Z}$. At $m_{Z}$, we have
the value $(\alpha_{em})^{-1}=128.8$ (Consoli, Jegerlehner,
Hollik, 1989, $[12]$), and we find, using at first
$\lambda_{Z}=1$, the value :
\begin{equation}m_{Z}=44.68GeV\end{equation}
Clearly, compared to the observed $m_{Z}=91.1887GeV$, we find that
the coefficient $\lambda_{Z}=2.057$, which is equal to $2$ with a
relative uncertainty of $2.9\times10^{-2}$, itself of the order of
$(\Omega-1)$. We thus see that our simplest formula gives us
unexpected good results, and seems to be able to probe the
structure of the $Z$ particle, predicting that this neutral
particle is made of two particles, probably of respectively
positive and negative electric charges.
\subsection{Strong interactions}
We know that the strong interactions are asymptotically free and
that the strong coupling has the following expression :
\begin{equation}\alpha_{s}(\mu)
=\frac{1}{4\pi\beta_{0}\ln(\mu^{2}/\Lambda_{QCD}^{2})}\left[1-\frac{\beta_{1}}{\beta_{0}^{2}}\frac{\ln(\ln(\mu^{2}/\Lambda_{QCD}^{2}))}{\ln(\mu^{2}/\Lambda_{QCD}^{2})}\right]
\end{equation}
(see for example $[13]$, and see references therein). So, if we
can compute the scale parameter $\Lambda_{QCD}$ with the help of
$\kappa_{0}$, the former formula unifies the coupling constant of
strong interactions and gravity. The value of $\Lambda_{QCD}$ is
determined by experiments, once specified the subtraction scheme
and the number of active flavors. For example
$\Lambda_{\overline{MS}}^{N_{f}=5}=217\pm24MeV$. If we compare
this mass to the power $\kappa_{0}^{-2/3}$, we find :
\begin{equation}\Lambda_{\overline{MS}}^{N_{f}=5}\approx30(\kappa_{0})^{-2/3}(\Omega^{1/2}H)\end{equation}
It appears clearly that $\kappa_{0}$ by itself governs the order
of magnitude of all parameters of the standard model of particles,
and that these parameters should be completely calculated using
only $\kappa_{0}$.
\section{When GUT unify all four interactions}
\subsection{De Broglie's dimensionless spacetime}
\subsubsection{Unifying all scales in the universe} Equation (3.34)
\begin{equation}G=\frac{1}{192\pi}\frac{\kappa_{0}^{2}}{\Omega H^{2}}\end{equation}
is exactly the type of formulas which unifies the smallest scales
and the largest scales of the universe. Using (3.34) with the
principle of constancy of the intensity of the gravitational
interaction, we arrive at the conclusion that the masses are
proportional to
\begin{equation}m\sim\Omega^{1/2}H\end{equation}
which opens the possibility of the complete calculation of the
masses of the fundamental particles using cosmological parameters.
There is an important remark to make here : formula (10.1) will be
proved completely in $[3]$ using the following hypothesis : the
calculations necessary to retrieve the abundances of the elements
in the early universe are the same in our quantum model and in the
standard cosmological model. Thus, the observations of the
abundances of the elements in the universe, turn to be an
experimental test of theories of unification. Nevertheless, to
compare the standard and quantum cosmological models, we will have
to REDEFINE SPACE AND TIME, and especially the running of time. To
achieve this goal, we will stick essentially to De Broglie's
analysis of a mass of a particle (and how he used this mass as a
clock). For this reason, we have called this new spacetime De
Broglie's dimensionless spacetime. Indeed, this space and time are
completely dimensionless.
\subsubsection{Unifying interactions}
We now turn to the interpretation of formula (6.6) :
\begin{equation}\kappa_{0}=\exp\left[-\frac{1}{\alpha_{em}}\right]\end{equation}
Does there exist in physics such kind of law, and what can we
deduce from this? Clearly this law is instanton-like, or also
tunnelling-like. If we note $g$ the coupling constant of a
non-abelian Yang-Mills gauge theory, a self-dual solution to the
euclidian Yang-Mills theory, an instanton, has in the path
integral the contribution :
\begin{equation}e^{-S}=e^{-8\pi^{2}/g^{2}}=e^{-2\pi/\alpha}\end{equation}
where $\alpha=g^{2}/4\pi$ (the instanton is only one possibility,
we could say as well that we should try to show that gravity is a
quantum tunnelling effect of electromagnetism). The behavior in
(10.4) is analog to the behavior of the gravitational coupling in
equation (10.3). We deduce that gravity is a nonperturbative relic
of electromagnetism, and eventually of the other interactions.
Thus, the most probable hypothesis is that there is no graviton in
the universe.
\subsubsection{Three new masses}
We saw that when $H$ is used as a unit of energy, there are two
fundamental masses in the universe, the mass of the electron,
which is governed by $(\kappa_{0})^{-2/3}$ and the mass of the $Z$
particle, which is governed by $(\kappa_{0})^{-3/4}$. Thus there
should be another mass
\begin{equation}m=\frac{\sqrt{\alpha_{em}}}{2^{3/4}}(\kappa_{0})^{-1/2}(\Omega^{1/2}H)\end{equation}
Also, there should be the mass :
\begin{equation}m=\frac{\sqrt{\alpha_{em}}}{2^{3/2}}(\kappa_{0})^{-1/3}(\Omega^{1/2}H)\end{equation}
Eventually, we can also construct a mass with $\kappa_{0}^{-1/4}$.
The question is : is one of these masses of physical interest? The
first mass is $m=\sqrt{\alpha_{em}}\times6K$ ($K$ for Kelvin) and
seems fine to be the mass of the neutrino $\nu_{e}$. The question
is to know if the second, or the third mass, is, or is not, the
mass of the photon.
\section{The link with the Grand Unified Theories}
\subsubsection{Another formula}
The gravitational charge of the electron is, using (8.4) :
\begin{equation}\kappa_{0}\tilde{m}_{e}\approx\sqrt{\alpha_{em}}\kappa_{0}^{1/3}\end{equation}
We recall the relation $m=\tilde{m}(\Omega^{1/2}H)$. We have :
\begin{equation}\kappa_{0}=\exp\left[-1/\alpha_{em}\right]\end{equation}
We know that $1/\alpha_{em}=137.036$ and it appears that
$1/\alpha_{GUT}\approx45$ (see for example Ross, 1984, chapter 6,
Fig 6.1, $[14]$). These numerical values lead directly to
\begin{equation}\frac{1}{\alpha_{GUT}}\approx\frac{1}{3\alpha_{em}}\end{equation}
We thus deduce that the gravitational charge of the electron is
\begin{equation}\frac{\kappa_{0}\tilde{m}_{e}}{\sqrt{\alpha_{em}}}\approx\kappa_{0}^{1/3}\approx\exp\left[-\frac{1}{\alpha_{GUT}}\right]\end{equation}
\subsubsection{Grand Unified Theories}
The physicists who constructed the Grand Unified Theories (GUT)
started from an experimental fact : the three couplings of all
interactions except gravity begin to converge at a energy scale
$M_{X}$. (See the book Ross, 1984, $[14]$ and all references
therein). These theories are known to unify the three interactions
(except gravity), but to possess two weaknesses : first they leave
too arbitrary parameters and no interaction with gravity. On the
contrary, formula (10.10) proves that gravity is nothing else than
some yet to be determined nonperturbative quantum effect
(instanton, tunnelling, of ELSE!) of these Grand Unified Theories,
so these GUT entirely contain gravity. Furthermore, (10.10) proves
that, via this non perturbative correspondence (we mean the
logarithm or inversely the exponential in (10.10), the three
couplings converge also to the gravitational coupling (the mass of
the electron, or more generally the mass of the particles). It
then appears that these theories contain the necessary information
to compute the masses, and thus should at the end contain no
arbitrary parameter. In other words, (10.10) seems to prove that,
in fact, the Grand Unified Theories unify all four interactions.
\subsubsection{The GUT energy scale}
The value of the GUT energy scale is $M_{X}=2\times10^{15}GeV$.
The value for such a mass, using our principle : "the simplest
formula gives the mass", is :
\begin{equation}M=\sqrt{\alpha_{em}}\kappa_{0}^{-1}(\Omega^{1/2}H)\end{equation}
We recall that at this scale, $\alpha_{em}=\alpha_{GUT}$, thus we
find $M=7.4\times10^{16}GeV\approx37M_{X}$.
\section{Consequences for the nature of gravity}
\subsubsection{Classical spacetime}
We simply add that we always placed ourselves in a classical
spacetime. What has been thought to be a breakdown of spacetime at
Planck length is in our theory simply the radiuses of particles.
These radiuses cannot be smaller, because they are controlled by
$\kappa_{0}$. Thus, we simply use a classical spacetime, and
insert in Feynman integrals the physical cut-off determined by
$\kappa_{0}^{-1}$. In this classical spacetime, there can be two
cases. We see that formula (10.10) links the mass of the electron
with what seems to be euclidian instantons. So there are two cases
for gravity : it can be a purely quantum relic (first case)
(nonperturbative correction) to the $SU(5)$ Yang-Mills theory (or
$SO(10)$, which seems to take into account the masses of the
neutrinos), or it can be a nonperturbative phenomenon which still
needs to be quantized (second case).
\subsubsection{The status of Einstein's theory (first case)}
However, there still is a problem with Einstein's theory. We
believe that each non abelian theory possesses its real,
observable nonperturbative (only attractive) purely quantum
effects (first case). In the case of $SU(5)$, these are gravity.
So, what is now the status of Einstein's theory? Clearly,
Einstein's theory is an effective theory which gives account for
this purely quantum effects, when these effects are seen at the
classical scale (in this case, there is no classical gravity,
gravity is only a PURELY QUANTUM RELIC). Then, from the classical
point of view, there is no gravity. However, if Einstein's gravity
is just an effective theory to describe at the classical scale
what we see of a purely quantum phenomena, THIS GRAVITY SHOULD NOT
BE QUANTIZED BY THE USUAL FORMALISM AVAILABLE FOR THE OTHER
INTERACTIONS. Simply because we cannot quantize something which is
already a purely quantum correction. Indeed, we quantize a
interaction to determine the deviations from the classical theory.
In the hypothesis of emergent gravity, gravity is already one of
these deviations (what we meant by : there is no classical
gravity). Beyond emergent gravity, the gravity we have been led to
a Poincare's gravity. In one of his works, Poincare emitted the
hypothesis that gravity simply did not exist as an interaction,
but was only a relic of electromagnetism. To explain this, he gave
the example : if the proton had an electric charge slightly
different from minus the charge of the electron, the planets would
have a slight global excess of electric charge and
electromagnetism could display, at their scale, a phenomenon
called gravity, which had no existence by itself, but was only a
relic of electromagnetism. In the example of Poincare, of course
we obtain a repulsive gravity, and we do not know if he has been
imagined, by this procedure, attractive gravities. Thus, formula
(10.10) proves, in our opinion, that we have arrived at a
Poincare's gravity : a (quantum) relic of the other interactions,
which happens to have effects that can be seen at the classical
level. To give an example, we consider the no-go theorem of
Weinberg-Witten which asserts that "an interacting graviton cannot
emerge from an ordinary quantum field theory in the same
spacetime". With our Poincare-like gravity, we can stay in four
dimensions, because we do not need to avoid Weinberg-Witten
theorem : what our analysis makes us believe at this stage, is
THAT THERE IS NO GRAVITON AT ALL. (we are aware that this
possibility has already been considered as an eventual hypothesis,
but we assert that formula (10.10) transforms it in THE SIMPLEST
AND MOST PHYSICAL CONJECTURE, SUGGESTED BY EXPERIMENT itself,
taking into account our present knowledge). What would mean
anyway, a boson for inertia, whereas at the quantum level, we
cannot define inertia for particles (because of their wavy nature,
and because inertia can only be defined for classical bodies). In
a future work, we will try to treat this problem and particularly
to discuss the wave/particle duality problem, which is at the
center of this question.
\subsubsection{The status of Einstein's theory (second case)}
If, and we now place ourselves in the second case, gravity is now
a nonperturbative euclidian-instanton-like phenomenon coming from
the $SU(5)$ Yang-Mills theory, we mean a phenomenon which has its
classical and quantized regimes, THERE IS NO GRAVITON EITHER.
Indeed, independently on how we can retrieve Einstein's theory
from this instanton-like phenomenon, the quantization of the
theory can be made by the rules of quantization of these supposed
instantons and not by the rules of quantization of general
relativity. So we arrive at the same conclusion : WE SHOULD NOT
TRY TO QUANTIZE GENERAL RELATIVITY (which still is an effective
theory). This idea is confirmed by our first hypothesis of the
quantum equation of gravity : a tensorial equation can take into
account quantum features of the theory. So even in the second case
which we are now studying, it is possible that general relativity
not only takes into account the classical behavior of our
classical instanton phenomenon, but also a part of (if not all)
the quantum side of this instanton phenomenon.
\subsubsection{A first try}
We now try to see how euclidian instanton effects can be put by
hand in the usual equations, written in Minkowskian spacetime.
Starting with the gauge theory
\begin{equation}S=\int F_{\mu\nu}F^{\mu\nu}d^{4}x\end{equation}
we add to this lagrangian the topological :
\begin{equation}S=\int F_{\mu\nu}F^{\mu\nu}d^{4}x+i\theta\int
F_{\mu\nu}\tilde{F}^{\mu\nu}d^{4}x\end{equation} We should obtain
our Poincare's gravity, because in the path integral,
$e^{iS/\hbar}$ will display the factor which is on the second hand
of equation (10.10). We clearly are aware that this first
solution, with no change, may lead to inconsistencies, but
starting from it, we should be able to improve this first try.
\subsubsection{Unification unifies the contingent and the fundamental}
Unification unified the smallest and largest scales of the
universe. However, it unifies now the most contingent and the most
fundamental : $\kappa_{0}$, and gravity, appears to be generated
by some relic which is a quantum correction of the three other
forces : it is the most contingent, it is so contingent that it
does not exist by itself (classically). And here is $\kappa_{0}$,
the most fundamental constant in physics, which governs everything
in the universe. \subsection{A few remarks} We see that up to
constants of order unity, the exponents of $\kappa_{0}$ determine
all quantities in the universe. The temperature of the cosmic
background radiation is $T=2.73K$ approximated by
\begin{equation}T=(\kappa_{0})^{-1/2}(\Omega^{1/2}H)=10.09K\end{equation}
The total entropy of the universe, the total number of photons are
governed by the exponent $(\kappa_{0})^{-3/2}$, the ratio eta of
the total number of baryons to the total number of photons is
governed by the ratio $(\kappa_{0})^{1/6}$, the value of $\theta$
in the strong CP problem could be $(\kappa_{0})^{1/6}$, coming
from a yet unknown procedure of unification, the total mass of the
universe $\tilde{M}$ is governed by $(\kappa_{0})^{-2}$.
\section{The mass gap problem}
The mass gap problem, presented by A. Jaffe and E. Witten (Jaffe,
Witten, 2000, $[15]$ as a millennium problem of the Clay
Mathematics Institute, corresponds to the following question : is
there, from the side of gauge theories, a mechanism that provides
us with an energy gap : "there must be some constant $\Delta>0$,
such that every excitation of the vacuum has energy at least
$\Delta$". And they add : "Since the vacuum vector $\Omega$ is
Poincare invariant, it is an eigenstate with zero energy, namely
$\widehat{H}\Omega=0$". The problem is to establish the existence
of a non-trivial quantum Yang-Mills theory that exhibits a mass
gap, the existence including definite axiomatic properties.
Furthermore, the supremum of such $\Delta$ is the mass $m$, and it
still has to be proved that $m<+\infty$. Of course the relations
that we proved concerning the masses of the fundamental particles
give an negative answer to this problem. Such a condition
$m\geq\Delta>0$, from the side of gauge theories, even in the case
in which it is possible mathematically, would ruin the whole
Yang-Mills method : it would be in contradiction with the fact
that the masses tend to zero since there are proportional to some
function of $\dot{a}$, multiplied by the factor $1/a$ which tends
to zero, whereas $\dot{a}$ varies only very little compared to
$a$. So, we turn to another problem which is : in the hypothesis
$m=\tilde{m}H$, try to establish the mass gap for $\tilde{m}$,
with the existence of a $\tilde{\Delta}$, defined by
$\Delta=\tilde{\Delta}H$. However, this problem also has a simple
solution, which is yes, and which is available in any case, that
is for any Yang-Mills theory. Why this? Because any quantum
Yang-Mills theory, as any quantum theory, verifies the relations
of incertitude of Heisenberg, and we shall see that in this case,
there is always a mass gap $\Delta$. Indeed, in our context,
equation (5.2) is equivalent to $\Delta=H$ and thus to
$\tilde{\Delta}=1$. We recall that the principle of the proof : we
use Heisenberg's incertitude relations to find a mass gap
depending on the interval of time of the observation, and we add
that this interval of time is necessarily less that the age of the
universe.
\subsection{The necessarily breaking of Poincare invariance}
The quantum equation of gravity does not satisfy anymore the
property of general covariance, and is not even Poincare
invariant. We explain now why this is a necessary condition on
every theory of quantum gravity. In fact, Poincare invariance, and
general covariance, will be lost once we renormalize the
gravitational constant. Indeed, the equations of general
relativity are
\begin{equation}R_{ik}-\frac{1}{2}R=8\pi GT_{ik}\end{equation}
If we stick to these equations, and try to guess to what changes
would lead quantum corrections, we return directly to our ideas
for the construction of the quantum equation of gravity. Using
renormalization theory, we deduce that quantum corrections lead to
the fact that the couplings run with energy. The equation is
tensorial, but in any tensorial equation, the constants also have
to be tensors : the constants are constant functions of space-time
variables, that have to keep their value in any change of
coordinates. If, after the first order corrections, $G$ depends on
energy, and since energy is not Poincare invariant, the Poincare
invariance is automatically broken, as we saw in $[3]$. However,
we also saw in $[3]$ that the vacuum is obtained when the right
hand side of a tensorial equation of gravity vanishes, in which
case the gravitational coupling disappears, leading to the fact
that all tensorial equations are equivalent. We used this property
to deduce that quantum gravity and general relativity are
equivalent in vacuum. Thus, in vacuum, the quantum equation of
gravity is Poincare invariant again, and even generally covariant.
Also, the vacuum of the equation of quantum gravity is zero. In
this equation, $\Lambda$ does not represent the energy of the
vacuum, it is a term another term representing (exactly : supposed
to mimic) the nonperturbative corrections to Einstein's theory.
However, we have the relation $\Omega_{\Lambda}=\Omega_{TOT}/2$.
If we note $\Omega_{\epsilon}$ the part of $\Omega_{TOT}$
corresponding to real matter, we have
$\Omega_{TOT}=\Omega_{\epsilon}+\Omega_{\Lambda}$ and thus :
$\Omega_{\Lambda}=\Omega_{\epsilon}$. The vacuum is characterized
by the relation $\Omega_{\epsilon}\rightarrow0$, which gives
straightforwardly $\Omega_{\Lambda}=0$.
\subsection{A negative answer to the mass gap problem}
We now return to our problem of finding a mass gap $\Delta$. The
masses can be written :
\begin{equation}m=\tilde{m}f(\dot{a})\frac{1}{a}\end{equation}
with $\tilde{m}$ a dimensionless constant mass, and $f(\dot{a})$
an undetermined function of $\dot{a}$. However, the mass gap can
be asked to $m$, in which case we can prove that the answer is no,
or can be asked to $\tilde{m}$, in which case we can prove the
answer is yes. We now ask the question to $m$. We place ourselves
in the case $\theta(a)=\theta_{0}a^{2}$. This case implies an
equation for $a$ which is (17.5) of $[1]$ :
\begin{equation}\dot{a}=\frac{1}{\sqrt{2\theta_{0}}}\sqrt{\ln(a/a_{0})}\end{equation}
We know from (14.7) that in any case, there exists $C>0$ and
constant, such that :
\begin{equation}m\leq C\frac{\dot{a}^{4}}{a}\sim\frac{\ln^{2}(a/a_{0})}{a}\end{equation}
So nothing can prevent this mass from tending to $0$ when
$a\rightarrow+\infty$, and there can be no mechanism, coming from
gauge theories, which keeps the mass :
\begin{equation}m\geq\Delta>0\end{equation}
This mechanism would be in contradiction with the forever
expansion of the universe. Indeed, the quantum equation of gravity
implies $\ddot{a}>0$, and the existence of a minimum of $a$ noted
$a_{0}$, for which $\dot{a}_{0}=0$. Such conditions on any
function $a(t)$ imply that $a(t)\rightarrow+\infty$ when
$t\rightarrow+\infty$. Indeed, $\dot{a}$ is strictly increasing
and starting from zero, becomes strictly positive, such that all
the tangent to the curve become strictly increasing. From this,
being convex, $a(t)\rightarrow+\infty$, because it is greater than
all its tangents.
\subsection{A positive answer to the mass gap problem}
If the question of the mass gap is asked to $\tilde{m}$, then the
answer is yes and comes simply from the Heisenberg's incertitude
relations. To distinguish any kind of positive energy from the
value zero, this energy must have a strictly positive value, of
course greater than the uncertainty coming from Heisenberg's
incertitude relations. These last relations thus imply that the
energy under consideration must take some value $\Delta$ such that
$\Delta\times T\geq h$, and $h$ is Planck's constant. Here, $T$ is
the value of the interval of time needed to observe the energy
$\Delta$. If we take for $T$ the greatest interval of time
possible, the age of the universe, we see that the Heisenberg
relation implies, using the condition $\hbar=c=1$,
$\Delta\geq1/t$, where now $t$ is the age of the universe.
Furthermore, we ask for this energy $\Delta$ to be written in the
dimensionless units of $\tilde{m}$. We thus are studying
$\tilde{\Delta}$, defined by :
\begin{equation}\Delta=\tilde{\Delta}f(\dot{a})\frac{1}{a}\end{equation}
In these dimensionless units, $\tilde{\Delta}$ is the smallest
energy possible, strictly greater than zero. Here $\tilde{\Delta}$
is a function depending on the cosmological time $t$, and we want
to know if it keeps values greater than some strictly positive
given value, during all times. In other words, we want to prove
that
\begin{equation}\frac{1}{\tilde{\Delta}}\leq Cte\end{equation}
We proved the condition $\Delta\geq1/t$, which implies that (15.7)
is equivalent to :
\begin{equation}\frac{1}{\tilde{\Delta}}=f(\dot{a})\frac{1}{a\Delta}\leq
f(\dot{a})\frac{t}{a}\leq Cte\end{equation} This condition can be
proved, provided we have in our quantum equation the right
function $\theta(a)$. From the mathematical point of view this is
enough, because finding only one function $\theta(a)$ is
equivalent to finding the mathematical mechanism, which was asked
for, in the formulation of the problem. To solve the problem from
a physical point of view, we have to take the particular function
$\theta(a)$, which has been proved the most physical in $[3]$ :
\begin{equation}\theta(a)=\theta_{0}a^{2}\end{equation}
In this case, we can apply (12.3) :
\begin{equation}\dot{a}=\frac{da}{dt}=\frac{1}{\sqrt{2\theta_{0}}}\sqrt{\ln(a/a_{0})}\end{equation}
where $a_{0}$ is the minimum value for the cosmological parameter
$a$. We saw earlier that there are two possibilities for the
universe before it reached the value $a=a_{0}$ : in one first
physical case, $a_{0}$ is only a minimum and the universe has
shrunk once to this value before expanding again, or in another
theoretical case, the values of $\theta$ are quantized, and before
the condition $a=a_{0}$, there has been a time of constant
$\theta$, with an exponential growth of $a$, with $a\rightarrow0$
for $t\rightarrow-\infty$. However, from the time when $a=a_{0}$,
the age of the universe has been
\begin{equation}t=\sqrt{2\theta_{0}}\int_{a_{0}}^{a}\frac{dx}{\sqrt{\ln(x/a_{0})}}\end{equation}
and we can take this time to be the longest interval of time for
the observation of a particle. Indeed, in case the universe has
grown to this value $a_{0}$ with an exponential law, $a$ comes
from zero, and there has been a time $t_{1}<t_{0}$, where the
radius where so near the value zero that the concept of particles
as we know them could not exist. So before time $t_{1}$, the
particle we study could not exist, and thus has not been observed.
From $t_{1}$ to $t_{0}$, there is only a finite interval of time.
Such a finite interval of time $(t_{0}-t_{1})$ counts for
$$f(\dot{a})\frac{t_{0}-t_{1}}{a}$$ in (12.8), and it is
easy to check that this term does not make arise any problem in
the following proof. The most interesting and physical case is
when $a_{0}$ is only a minimum of $a$. We proved in $[3]$ that in
this case, time before $a_{0}$, runs backwards and not forwards.
Indeed, the big bang, that is to say the universe at the value
$a_{0}$, does not possess a past from which the universe has
shrunk to $a_{0}$ and a future, into which the universe is growing
again. On the contrary this point has no past but two futures,
probably equivalent. In other words, and this shall be proved
carefully, if we could look through the big bang, we would not see
some kind of past of the big bang, or cause of the big bang, but
we would see one future, acting as a cause of ourselves, in a twin
universe in which the effect is always before the cause. This is
clearly the logical solution for the existence of the universe :
there are twin universes, each of them being the cause of the
existence of the other. In this case, the greatest interval of
time that can be considered is the interval of time from $a_{0}$
until now. We make the change of variables $x=a_{0}y$ in the
integral (12.11), and our condition becomes :
\begin{equation}f(\dot{a})\frac{t}{a}=\frac{f(\dot{a})}{a}\sqrt{2\theta_{0}a_{0}^{2}}\int_{1}^{a/a_{0}}\frac{dy}{\sqrt{\ln
y}}\leq Cte\end{equation} There is no problem for the integral at
the values $y\approx1$, because :
$$\int_{1}^{2}\frac{dy}{\sqrt{\ln y}}<+\infty$$
For the smallest values of $a$, the expression in (12.12) is less
or equal than a constant, because $a$ does not tend to zero but to
a strictly positive value $a_{0}$. Furthermore, when $a\rightarrow
a_{0}$, we have $\dot{a}\rightarrow0$. The worst behavior which we
consider for $f(\dot{a})$ is the case $m\sim\Omega^{1/2}H$,
equivalent to $f(\dot{a})=\sqrt{1+\dot{a}^{2}}$. When
$\dot{a}\rightarrow0$, we find $\Omega^{1/2}H\sim1/a_{0}$, which
is a constant, so $f(\dot{a})t/a$ has no problem in this limit.
For $a\rightarrow+\infty$, we compute an equivalent of the
integral. If $X$ is a real variable, the derivative of
$X/\sqrt{\ln X}$ is :
\begin{equation}\left(\frac{X}{\sqrt{\ln X}}\right)'=\frac{1}{\sqrt{\ln
X}}-\frac{1}{2(\ln X)^{3/2}}\sim\frac{1}{\sqrt{\ln
X}}\end{equation} Here the symbol $f(X)\sim g(X)$ means that the
ratio $f(X)/g(X)$ tends to $1$ when $X\rightarrow+\infty$. We
integrate this relation, finding :
\begin{equation}\int_{1}^{X}\frac{dy}{\sqrt{\ln y}}\sim\int_{e}^{X}\frac{dy}{\sqrt{\ln y}}\sim\int_{e}^{X}\left(\frac{y}{\sqrt{\ln
y}}\right)'dy\sim\frac{X}{\sqrt{\ln X}}\end{equation} We thus find
that, for $a\rightarrow+\infty$ :
\begin{equation}\frac{f(\dot{a})t}{a}\sim\frac{f(\dot{a})\sqrt{2\theta_{0}}a_{0}}{a}\int_{1}^{a/a_{0}}\frac{dy}{\sqrt{\ln
y}}\sim
f(\dot{a})\frac{\sqrt{2\theta_{0}}}{\sqrt{\ln(a/a_{0}})}\end{equation}
We also have
$$\dot{a}=\frac{1}{\sqrt{2\theta_{0}}}\sqrt{\ln(a/a_{0})}$$
and we obtain :
\begin{equation}\frac{f(\dot{a})t}{a}\sim\frac{f(\dot{a})}{\dot{a}}\end{equation}
We first make one important remark : equation (12.16) is the key
equation to prove the mass gap problem, and is at the same time
equivalent to the following relation :
$$\frac{t}{a}\sim\frac{1}{\dot{a}}$$
or in other words, (12.16) is equivalent to
\begin{equation}t\sim\frac{1}{H}\end{equation}
Equation (12.17), equivalent to (12.16), itself equivalent to the
solution of the mass gap problem, has been experimentally observed
(Bennett and al. 2003, $[5]$). It is also really important to
notice that the result (12.16) is independent of the value of
$\theta_{0}$. Indeed, physically, we would like to see integers
emerging in the theory, because we preview that unification will
be a theory involving only integers. To obtain these integers, we
would like to find $\tilde{\Delta}=1$, or at least find that the
value of $\tilde{\Delta}$ does not involve any parameter of the
equation, like $\theta_{0}$. The condition solving the mass gap
problem is then
\begin{equation}\frac{f(\dot{a})}{\dot{a}}\leq Cte\end{equation}
Now, if we note $a_{1}$ and $\dot{a}_{1}$ the present values
respectively of the radius of the universe and of its time
derivative, the complete relation for $\dot{a}$ is :
\begin{equation}\dot{a}^{2}=\dot{a}_{1}^{2}+\frac{1}{2\theta_{0}}\ln(a/a_{1})\end{equation}
We saw that the value of $\dot{a}_{1}$ should be around unity, at
most $20$, the greatest value considered in $[3]$ being $7.07$.
Now, if we just look at the behavior of $\dot{a}$, we see that
when the universe will be $(\kappa_{0})^{-1}$ times bigger that it
is today, we recall that for this value, the radius of this future
universe will be to the present radius what the present radius is
today to the Planck length, for this value we said, and for
$\dot{a}_{1}\approx20$ the value of $\dot{a}$ will only be around
$23$. Indeed, we proved in $[2]$ that $2\theta_{0}=1$. We thus
understand that the only possible values of $\dot{a}$ are around
unity, and since $f(\dot{a})$ is a continuous function, (12.18) is
satisfied since $\dot{a}$ stays around unity. We add that the
value of $(\kappa_{0})^{-1}$ has been considered as the greatest
value ever imaginable in our world. This means that in fact the
ratio $a/a_{1}$, for one yet unknown reason, has no chance to go
beyond this value. We should imagine that to one point, when this
ratio becomes too big, the laws of physics as we elaborated them
should breakdown completely. We should probably have a universe
with only particles with vanishing masses.
\subsection{Another proof of the existence of a mass gap}
The former proof is the most general : it uses no further
hypothesis on $f(\dot{a})$. We can also prove the existence of the
mass gap, that is to say (12.18), by turning to experiment to
consider the most physical function $f(\dot{a})$, or even to the
quantum cosmological model under consideration, plus find a
physical principle that will permit us to compute $f(\dot{a})$. We
saw earlier that we could find the best estimates for the masses
of the fundamental particles when we supposed that these were
proportional to
\begin{equation}m\sim\Omega^{1/2}H=\frac{\sqrt{1+\dot{a}^{2}}}{a}\end{equation}
or eventually to
\begin{equation}m\sim H=\frac{\dot{a}}{a}\end{equation}
For example, the computation of the mass of the electron is in
perfect agreement with (12.20) but does not rule out completely
(12.21). In these cases, we can compute $f(\dot{a})$ : (12.20)
leads to the relation :
\begin{equation}f(\dot{a})=\sqrt{1+\dot{a}^{2}}\end{equation} and
we have :
\begin{equation}\frac{f(\dot{a})}{\dot{a}}=\frac{\sqrt{1+\dot{a}^{2}}}{\dot{a}}\end{equation}
This is a decreasing function of $\dot{a}$, and also a decreasing
function of time since $\ddot{a}>0$. So $f(\dot{a})/\dot{a}$
decreases to $1$ as $t\rightarrow+\infty$ and we now find the
exact relation for $\tilde{\Delta}$ :
\begin{equation}\tilde{\Delta}=1\end{equation}
Equation (12.21) is even simpler and in this case
\begin{equation}\frac{f(\dot{a})}{\dot{a}}=1\end{equation}
and equation (12.24) is still satisfied. We can also turn to a
physical principle to be associated with our quantum cosmological
model. We can suppose that the total number of baryons is constant
in the universe, which leads to the relation :
\begin{equation}m\sim\frac{(1+\dot{a}^{2})^{2}}{a}\end{equation}
which equation (14.7) of $[3]$ p 209. This behavior of $m$ is too
bad, and (12.26) cannot improve the general proof of section 12.3.
On the contrary, if we use our quantum cosmological model with the
additional hypothesis of the constancy of the intensity of the
gravitational interaction, we use equation (14.3) of $[3]$ p 209
to deduce that Newton's constant is proportional to
\begin{equation}G\sim\frac{\kappa_{0}}{\sqrt{\epsilon}}\sim\frac{1}{\Omega
H^{2}}\end{equation} The constancy of the gravitational charge is
: $$G^{1/2}m\sim Cte$$ From this equation, we deduce directly
(12.20) and conclude.
\subsection{A third proof of the mass gap problem}
There is finally a special case of our class of quantum
cosmological models for which no additional principle is necessary
to compute $f(\dot{a})$, and directly conclude to the existence of
a mass gap. This is the limit case $\theta\rightarrow+\infty$.
Equation (15.18) of $[1]$ reads :
$$\frac{\ddot{a}}{a}=\frac{1}{4\theta(a)}$$
In the case $\theta\rightarrow+\infty$, the last relation becomes
$\ddot{a}=0$, and thus $\dot{a}=\lambda$, where $\lambda$ is a
constant. We integrate this relation and find $a(t)=\lambda t$,
since we still can decide to choose $t=0$ for $a=0$. We thus
obtain :
\begin{equation}f(\dot{a})\frac{t}{a}=\frac{f(\lambda)}{\lambda}=Cte\end{equation}
and (12.12) of the present article is satisfied, which proves once
again the existence of a mass gap.
\subsection{A simplified picture of the proof}
We give a simplified proof, in the picture in which the masses are
proportional to Hubble's constant :
\begin{equation}m=\tilde{m}H\end{equation}
From the observations of Bennett and al. 2003 $[5]$, or from our
theoretical analysis in $[1]$, section 17, we know that the age of
the universe is :
\begin{equation}t\approx\frac{1}{H}\end{equation}
The smallest energy $\Delta$ that can be distinguished from zero
is :
\begin{equation}\Delta\approx\frac{1}{t}\approx H\end{equation}
Using now :
\begin{equation}\Delta=\tilde{\Delta}H\end{equation}
we find again the right formula :
\begin{equation}\tilde{\Delta}\approx1\end{equation}
In fact, we see further that this condition implies that in the
relation : \begin{equation}m=\tilde{m}H\end{equation} the
incertitude on the value of $\tilde{m}$ is around unity, so there
is no value of this variable that can be distinguished from its
nearest integer. This way, we can suppose that $\tilde{m}$ only
takes integer values. This formulation in fact is equivalent to
the condition $\tilde{\Delta}=1$.

\vspace{10mm} Email address : cristobal.real@hotmail.fr


\vspace{10mm}



\begin{thebibliography}{99}


\bibitem{Re1} C. R\'{e}al, {\em Tensorial Quantum Gravity and the Cosmological Constant Problem}, arXiv:0711.1441.

\bibitem{Re2} C. R\'{e}al, {\em Physical Unification and Masses}, in {\em Unification
and the Masses of the Fundamental Particles}, Editions Cristobal,
2007.

\bibitem{Re3} C. R\'{e}al, {\em Unification and the Masses of the
Fundamental Particles}, Editions Cristobal, 2007.

\bibitem{Lochak} G. Lochak, {\em The Equation of a Light Leptonic
Magnetic Monopole and its Experimental Aspects} Z. Naturforsch.
62a, 231-246 (2007).

\bibitem{Be} C. L. Bennett and al., {\em First Year Wilkinson Microwave
Anisotropy Probe (WMAP) Observations : Preliminary Maps and Basics
Results}, astro-ph/0302207, Astrophysical Journal.

\bibitem{DBro63} L. De Broglie, {\em La Thermodynamique de la particule
isolée}, Gauthier-Villars, 1963 .

\bibitem{Pe93} P. J. E. Peebles, {\em Principles of Physical
Cosmology} Princeton : Princeton University Press, 1993.

\bibitem{LL4} L. Landau, E. Lifchitz, {\em Physique Theorique
Electrodynamique Quantique; Quantum Electrodynamics} Ed. Librairie
du Globe; Editions Mir.

\bibitem{Yao} Yao and al., J.Phys. G33, 1, 2006.

\bibitem{MS} F. Mandl, G. Shaw, {\em Quantum Field Theory}, John
Wiley and Sons, 1984, reprinted in 1995.

\bibitem{Pi} A. Pich, {\em The Standard Model of Electroweak
Interactions} in {\em The Standard Model and Beyond}, Editions
Frontieres, 1995, p. 1-41.

\bibitem{CJH} M. Consoli, F. Jegerlehner and W. Hollik, {\em Z
Physics at LEP} CERN Yellow Report CERN 89-08, Vol. 1, CERN
Geneva, 1989, p.55.

\bibitem{YHM} K. Yagi, T. Hatsuda, Y. Miake, {\em Quark-Gluon
Plasma}, Cambridge Monographs on Particle Physics, Nuclear Physics
and Cosmology, 2005.

\bibitem{Ross} G. G. Ross, {\em Grand Unified Theories}, Frontiers
in Physics, 1984.

\bibitem{JW} A. Jaffe, E. Witten, {\em Quantum Yang-Mills Theory},
Millennium Prize Problem, Clay Mathematics Institute.


\end{thebibliography}
\end{document}